\documentclass[pdflatex,twoside,sn-mathphys-num]{sn-jnl}

\usepackage{float}
\usepackage{graphicx}%
\usepackage{multirow}%
\usepackage{amsmath,amssymb,amsfonts}%
\usepackage{amsthm}%
\usepackage{mathrsfs}%
\usepackage[title]{appendix}%
\usepackage{xcolor}%
\usepackage{textcomp}%
\usepackage{manyfoot}%
\usepackage{booktabs}%
\usepackage{algorithm}%
\usepackage{algorithmicx}%
\usepackage{algpseudocode}%
\usepackage{listings}%
\usepackage{setspace}
\usepackage{caption}
\usepackage[skip=3pt]{caption}  
\theoremstyle{thmstyleone}%
\theoremstyle{thmstyletwo}%

\theoremstyle{thmstylethree}%

\begin{document} 
\title[Article Title]{Temperature dependent ferroelectricity in strained KTaO$_3$ with machine learned force field}

\author[1,3]{\fnm{Yu} \sur{Zhu}}

\author[1]{\fnm{Luigi } \sur{Ranalli}}

\author[1,3]{\fnm{Taikang} \sur{Chen}}

\author*[3]{\fnm{Wei} \sur{Ren}}\email{renwei@shu.edu.cn}

\author*[1,2]{\fnm{Cesare} \sur{Franchini}}\email{cesare.franchini@univie.ac.at}

\affil[1]{\orgdiv{Faculty of Physics and Center for Computational Materials Science}, \orgname{University of Vienna},\city{ Vienna}, \postcode{1090}, \state{State}, \country{Austria}}

\affil[2]{\orgdiv{Department of Physics and Astronomy "Augusto Righi"
Alma Mater Studiorum}, \orgname{Università di Bologna}, \city{Bologna}, \postcode{40127}, \country{Italy}}

\affil[3]{\orgdiv{Department of Physics,Shanghai Key Laboratory of High Temperature Superconductors, International Centre of Quantum and Molecular Structures}, \orgname{Shanghai University}, \city{Shanghai}, \postcode{200444},  \country{China}}

\maketitle  
\section*{\centering Abstract}

\begin{spacing}{2} 
Ferroelectric materials are a class of dielectrics that exhibit spontaneous polarization which can be reversed under an external electric field. The emergence of ferroelectric order in incipient ferroelectrics is a topic of considerable interest from both fundamental and applied perspectives. Among the various strategies explored, strain engineering has been proven to be a powerful method for tuning ferroelectric polarization in materials. In the case of KTaO$_3$, first principles calculations have suggested that strain can drive a ferroelectric phase transition. 
In this study, we investigate the impact of in-plane uniaxial and biaxial strain, ranging from 0\% to 1\%, on pristine KTaO$_3$ to explore its potential for ferroelectricity induction via inversion symmetry breaking. By integrating density functional theory calculations with the stochastic self-consistent harmonic approximation assisted by on the fly machine learned force field, we obtain accurate structural information and dynamical properties under varying strain conditions while incorporating higher-order anharmonic effects. Employing the Berry phase method, we obtained the ferroelectric polarization of the strained structures over the entire temperature range up to 300 K.
 Our findings provide valuable insights into the role of strain in stabilizing ferroelectricity in KTaO$_3$, offering guidance for future experimental and theoretical studies on strain-engineered ferroelectric materials.

\vspace{1em}  
\noindent\textbf{Keywords:} ferroelectric transition, KTaO$_3$, anharmonic phonons, SSCHA
\end{spacing}

\newpage

\maketitle
\section{Introduction}\label{sec1}
\begin{figure}[H]
\vspace{1pt}
    \centering
\includegraphics[width=0.8\linewidth]{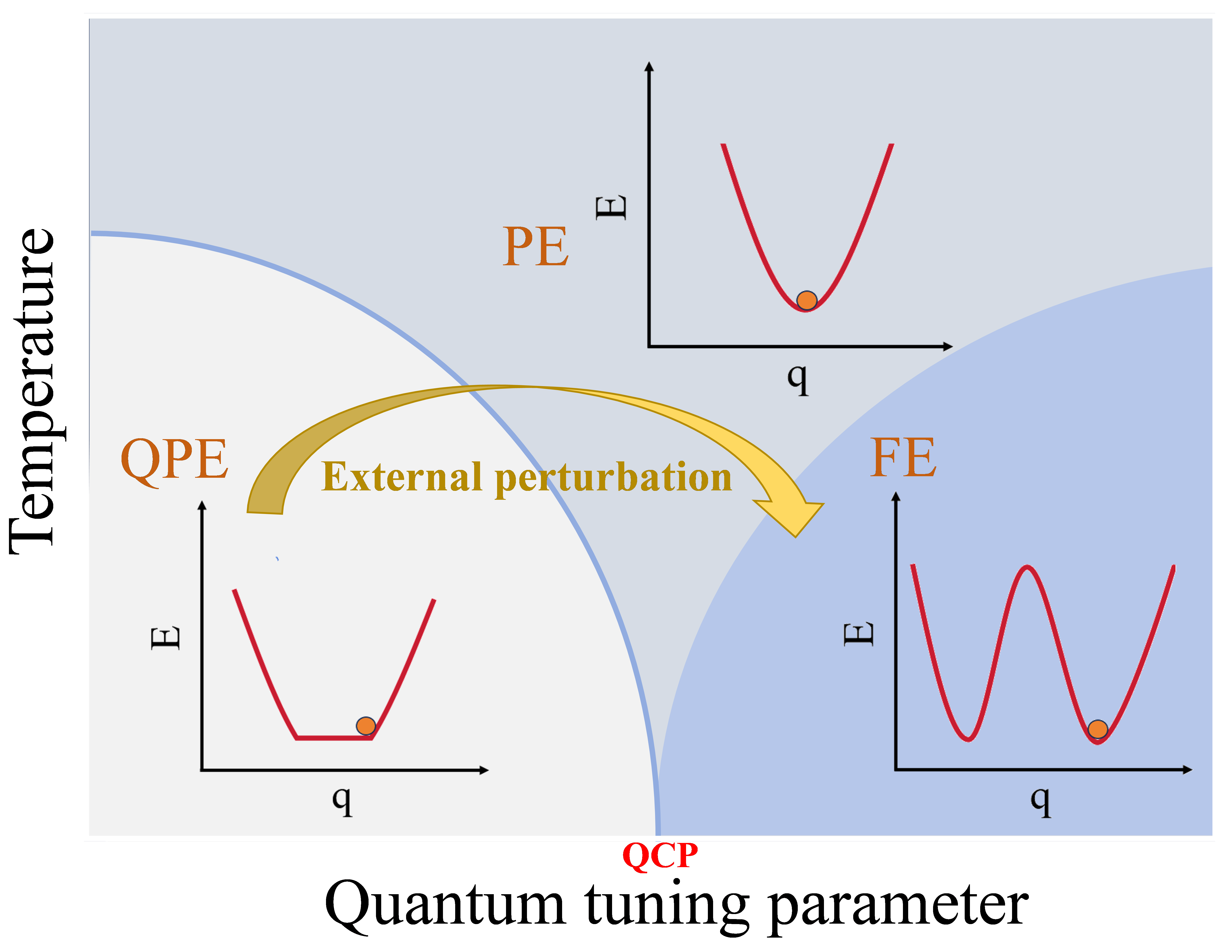}
    \caption{Qualitative phase diagram of the crossover among the ferroelectric (FE), paraelectric (PE) and quantum paraelectric (QPE) phases as the function of quantum
tuning parameter and temperature}
    \label{fig:QCP}
\end{figure}

\begin{spacing}{2}
Potassium tantalate (KTaO\(_3\)) is a prototypical quantum paraelectric, in which strong quantum fluctuations suppress the onset of ferroelectric order even at the lowest temperatures~\cite{muller1979srti,kvyatkovskii2001quantum}. It exhibits a temperature dependent transverse optical soft phonon mode whose frequency decreases upon cooling but never vanishes~\cite{cowley1962temperature,migoni1976origin}, leading to a pronounced deviation of the inverse dielectric constant from the classical Curie–Weiss law~\cite{fujishita2016quantum,rowley2014ferroelectric}. This soft mode saturation and associated dielectric anomaly place KTaO\(_3\) in close proximity to a ferroelectric quantum critical point (QCP)~\cite{rowley2014ferroelectric,sachdev1999quantum} in Figure~\ref{fig:QCP}.
Such proximity implies that small external perturbations—such as strain~\cite{haeni2004room,tyunina2010evidence}, hydrostatic pressure~\cite{samara1966pressure,fujii1987stress,uwe1976stress}, or isotopic substitution~\cite{hochli1977quantum,itoh1999ferroelectricity,itoh2000quantum}—can drive the system across the QCP and induce long range ferroelectricity, a hallmark of so-called incipient ferroelectrics. 

Over the past decades, numerous experiments have confirmed that the dielectric properties of KTaO\(_3\) is highly tunable~\cite{wemple1965some,ang2001dielectric,shirane1967temperature,haeni2004room}, yet the microscopic mechanisms governing the interplay between external perturbations and quantum fluctuations remain incompletely understood~\cite{souvatzis2009self,hellman2011lattice,errea2013first,tadano2015self,van2021quantum,zacharias2023anharmonic,esswein2023first,saha2025thermodynamic}.
From a theoretical standpoint, modeling quantum paraelectric from first principles is challenging. Conventional density functional theory (DFT), formulated at zero temperature, cannot capture quantum or thermal fluctuations and thus fails to describe the suppression of ferroelectricity in materials like KTaO\(_3\)~\cite{tao2016strain}. Path integral Monte Carlo simulations with effective Hamiltonian~\cite{zhong1996effect,iniguez2002first,akbarzadeh2004atomistic} and low-dimensional lattice nuclear Schrödinger equation~\cite{shin2021quantum,esswein2022ferroelectric} approaches have provided insights, but they often neglect mode coupling and are restricted to simplified models. More recently, the stochastic self-consistent harmonic approximation (SSCHA)~\cite{hooton1955li,monacelli2021stochastic,bianco2017second,errea2014anharmonic} has emerged as a powerful framework to treat anharmonic phonons nonperturbatively while fully including nuclear quantum and thermal effects, offering a route to qualitatively describe quantum criticality in incipient ferroelectrics~\cite{ranalli2023temperature,verdi2023quantum,bernhardt2024ferroelectric,ranalli2024electron,schmidt2025machine}. However, the requirement of large supercells for accurate phonon research makes the computational cost prohibitively high. Machine learning algorithms have been demonstrated to effectively accelerate such simulations while maintaining accuracy, thus providing a promising strategy to overcome this technical limitation~\cite{schmidt2019recent,deringer2019machine,unke2021machine}.

Despite these advances, a key open question remains: how do quantum fluctuations compete with strain-induced ferroelectric order in KTaO\(_3\). Understanding this competition is essential to determine whether long range ferroelectricity can be sustained in strained KTaO\(_3\) and how its temperature dependence evolves near the quantum critical regime. 
To address this question,we employ the SSCHA method ~\cite{hooton1955li,monacelli2021stochastic,bianco2017second,errea2014anharmonic}combined with an on the fly machine learned force field (MLFF) scheme \cite{jinnouchi2019fly,jinnouchi2019phase,jinnouchi2020descriptors} to investigate the competition between ferroelectric order and quantum fluctuation  in strained KTaO\(_3\) with first principles accuracy.  
\end{spacing}
\vspace{-3mm}
\section{Results}\label{sec2}

\begin{spacing}{2}

We apply in plane uniaxial
and biaxial strain ranging from 0\% to 1\% on pristine KTaO\textsubscript{3} to invesitigate its structural evolution and dielectric response over the temperature range of 0–300 K.The SSCHA+MLFF
workflow, introduced in the methodology section, enables us to obtain accurate
structural information under different strain conditions while incorporating
higher order anharmonic effects.

\begin{figure}[H]
\vspace{5pt}
    \centering
\includegraphics[width=\linewidth]{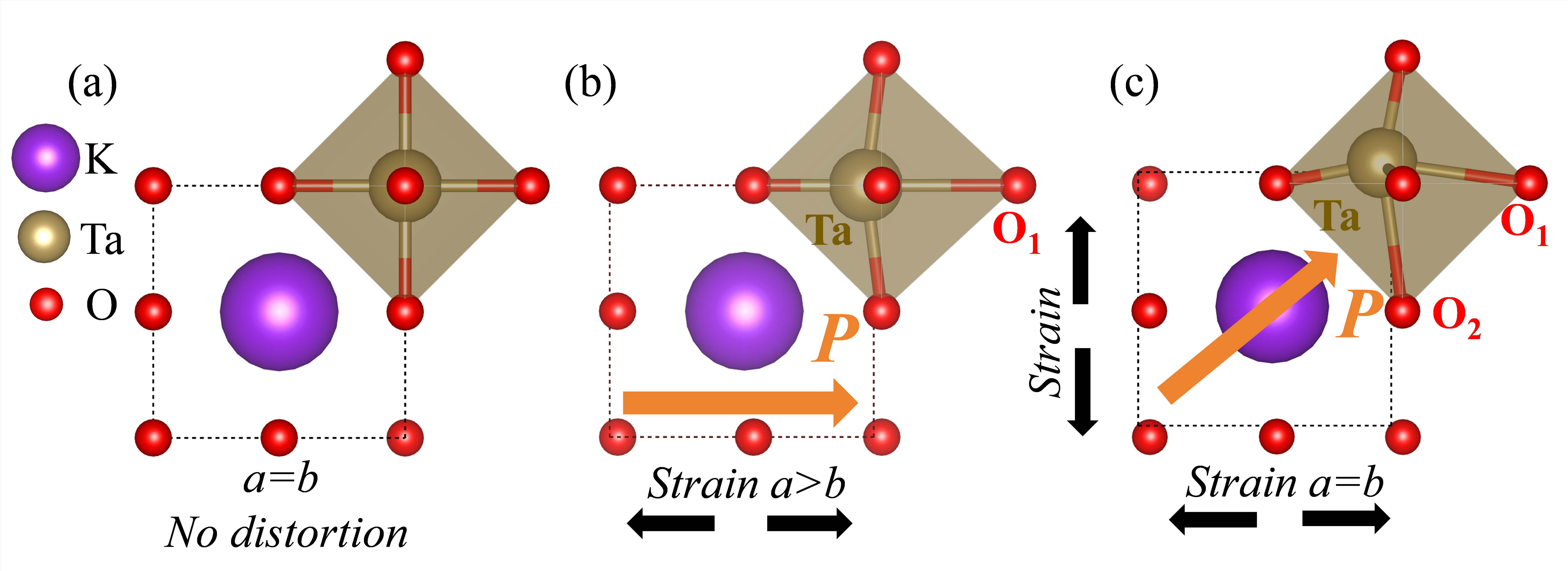}
    \caption{Geometric structure of (a) pristine KTaO\textsubscript{3}(b) uniaxial strained KTaO\textsubscript{3} and (c) biaxial strained KTaO\textsubscript{3}}
    \label{fig:str}
\end{figure}
\vspace{-15pt}
Regarding the structural details, as shown in Fig.~\ref{fig:str}, both uniaxial and biaxial strain  induce symmetry transformation and significant lattice distortions. The application of strain leads to structural transformation from cubic to tetragonal (uniaxial strain) or orthorhombic (biaxial strain) phase. Our calculations indicate that at low temperatures, KTaO\textsubscript{3} under uniaxial and biaxial strain adopts the space group P4mmm
 and Amm2, respectively. In contrast, pristine KTaO\textsubscript{3} belongs to the P4/mmm space group. Both P4mm and Amm2 exhibit lower symmetry than P4/mmm, with broken inversion symmetry, enabling ferroelectricity. And the Amm2 phase possesses even lower symmetry than P4mm phase, as the tetragonal fourfold rotational symmetry is broken.    

\begin{figure}[H]
    \centering
\vspace{5pt}    \includegraphics[width=1\linewidth]{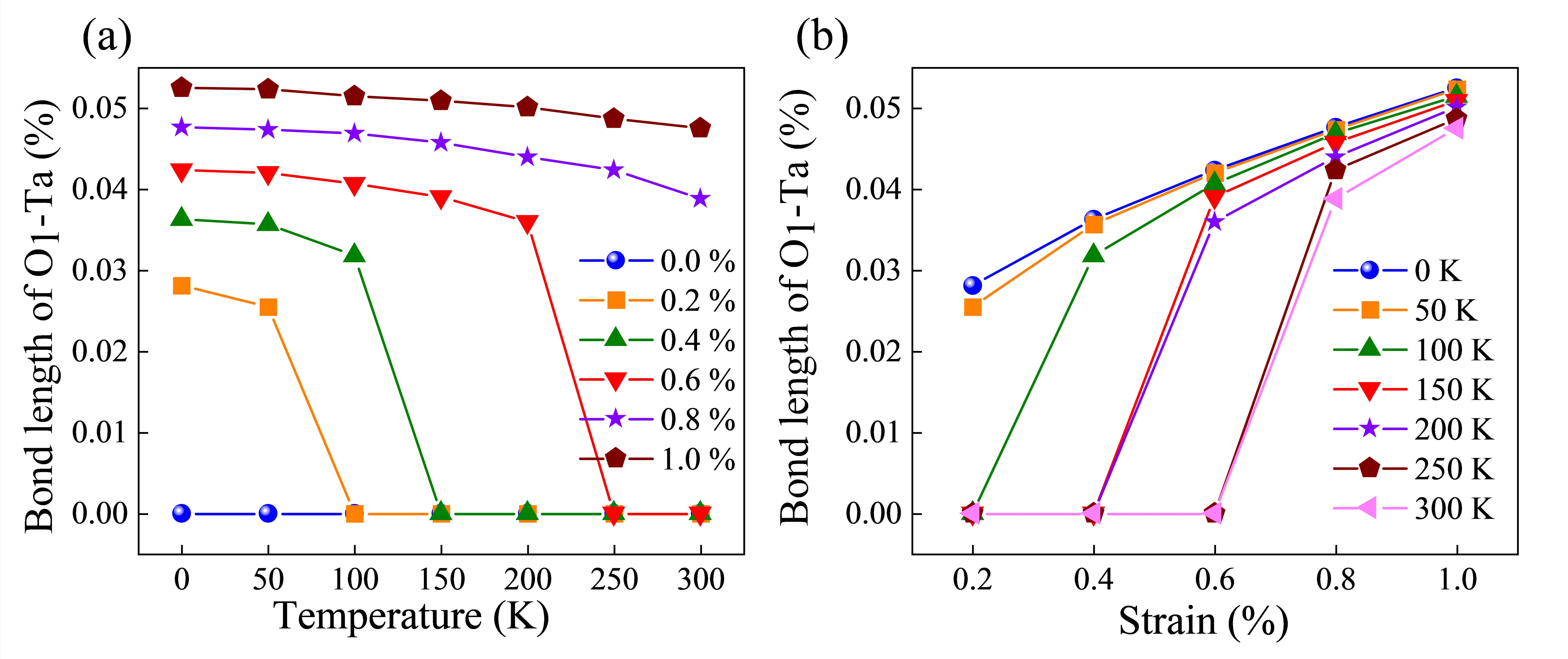}
    \caption{Variation of bond length O\textsubscript{1}-Ta for KTaO\textsubscript{3} under uniaxial strain and temperatures, only data points corresponding to strained structures that
exhibit distortion are included.}
    \label{fig:bond}
\end{figure}

The O\textsubscript{1}-Ta bond and O\textsubscript{2}-Ta bond extend along the a and b directions respectively, and their changes can represent structural distortions along these two directions. Therefore, to further illustrate the impact of strain and temperature on structural distortions in strained KTaO\textsubscript{3}, we fix the potassium atomic positions and analyze the variation of O\textsubscript{1}-Ta bond length and O\textsubscript{2}-Ta bond length (see in Fig.~\ref{fig:bond} and Fig. S1). These bond length variations serve as a key indicator of strain and temperature induced structural distortions.  It is evident that for a given strain, the variation in the O\textsubscript{1}-Ta bond length decreases with increasing temperature. This implies that the system exhibits reduced distortion at higher temperatures, progressively transitioning from a ferroelectric state to a paraelectric state. Under uniaxial strain, the space group evolves from P4mm to P4/mmm, and under biaxial strain, from Amm2 to P4/mmm. 
This transition is characterized by distinct Curie temperatures, marking the temperature at which the system loses its spontaneous polarization and restores higher symmetry. The system undergoes a symmetry transformation from a polar to a nonpolar space group.
Furthermore, the lowest temperature condition yields the most significant bond length variation across all strain levels, highlighting the strong temperature dependence of the strain-induced lattice distortions. 
 
\vspace{-15pt}
At a fixed temperature, the bond length variation increases as the strain magnitude increases, indicating a direct correlation between strain and structural distortion. Notably, the largest applied strain consistently exhibits the most pronounced bond length variation across the entire temperature range. The most pronounced distortion is observed under 1\% strain at 0~K, emphasizing the strong interplay between strain and temperature in governing the structural properties of KTaO\textsubscript{3}. 
Compared to uniaxial strained KTaO\textsubscript{3}, the only difference under biaxial strain is that structural distortions emerge along both the a and b directions. However, the magnitudes of distortion in these directions remain very similar. And the distortion magnitudes under uniaxial and biaxial strain are nearly identical.
The results underscore the critical role of both external strain and thermal effects in modulating the lattice distortions. Since the bond length variation is directly correlated with ferroelectric distortion, it provides valuable insight into the emergence of ferroelectric polarization, which will be discussed in detail in subsequent analysis.

\begin{figure}[H]
    \centering
\vspace{5pt}    \includegraphics[width=0.8\linewidth]{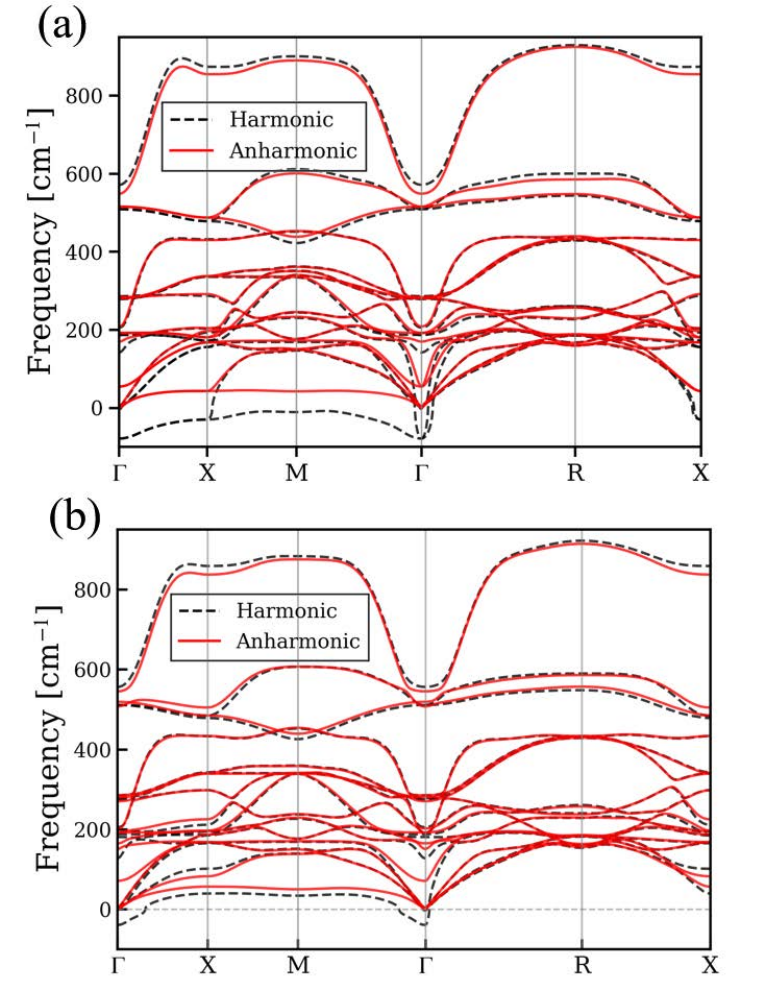}
    \caption{Phonon spectrum of (a) 1\% uniaxial strained KTaO\textsubscript{3} and (b) 1\% biaxial strained KTaO\textsubscript{3} at 0 K without (black dash line) and with (red line) anharmonic
effects.}
    \label{fig:Picture4}
\end{figure}
\vspace{-15pt}
For the dynamical properties, we obtained the phonon spectrums of both uniaxial strained KTaO\textsubscript{3} and biaxial strained KTaO\textsubscript{3} under 1\% strain and 0 K, as shown in Fig.~\ref{fig:Picture4}. 
The calculated data shows that, similar with pristine KTaO\textsubscript{3}~\cite{ranalli2023temperature}  including anharmonic terms can stabilize the spurious imaginary ferroelectric soft mode predicted by harmonic approximation.
Moreover, the other phonon branches remain unaffected by the anharmonic effects. This highlights the crucial role of anharmonicity, which significantly impacts the renormalized frequency of soft mode phonons. Ultimately, this interplay determines whether a material can cross the QCP into a ferroelectric phase. Neglecting anharmonicity can lead to severe misjudgments in predicting polarization dynamics and phase transition temperatures.

\end{spacing}
\vspace{-3mm}
\begin{spacing}{2}

For the quantum paraelectric phase KTaO\textsubscript{3}, the energy barrier for ferroelectricity is small compared to quantum fluctuations preventing the system from transitioning into ferroelectric state at any time. According to previous studies, the energy barrier of pristine KTaO\textsubscript{3} is approximately $0.306$~meV/atom~\cite{ranalli2023temperature}, while that of BaTiO\textsubscript{3} is around $4$~meV/atom~\cite{verdi2023quantum}. This comparison indicates that the energy barrier in KTaO\textsubscript{3} is significantly lower, allowing quantum fluctuations to readily overcome it and thereby stabilizing the quantum paraelectric phase. These results are obtained from DFT calculations using the Strongly Constrained
and Appropriately Normed (SCAN) functional\cite{sun2015strongly}, which we also employ for this study.

\begin{figure}[H]
\vspace{-12pt}
    \centering
    \includegraphics[width=\linewidth]{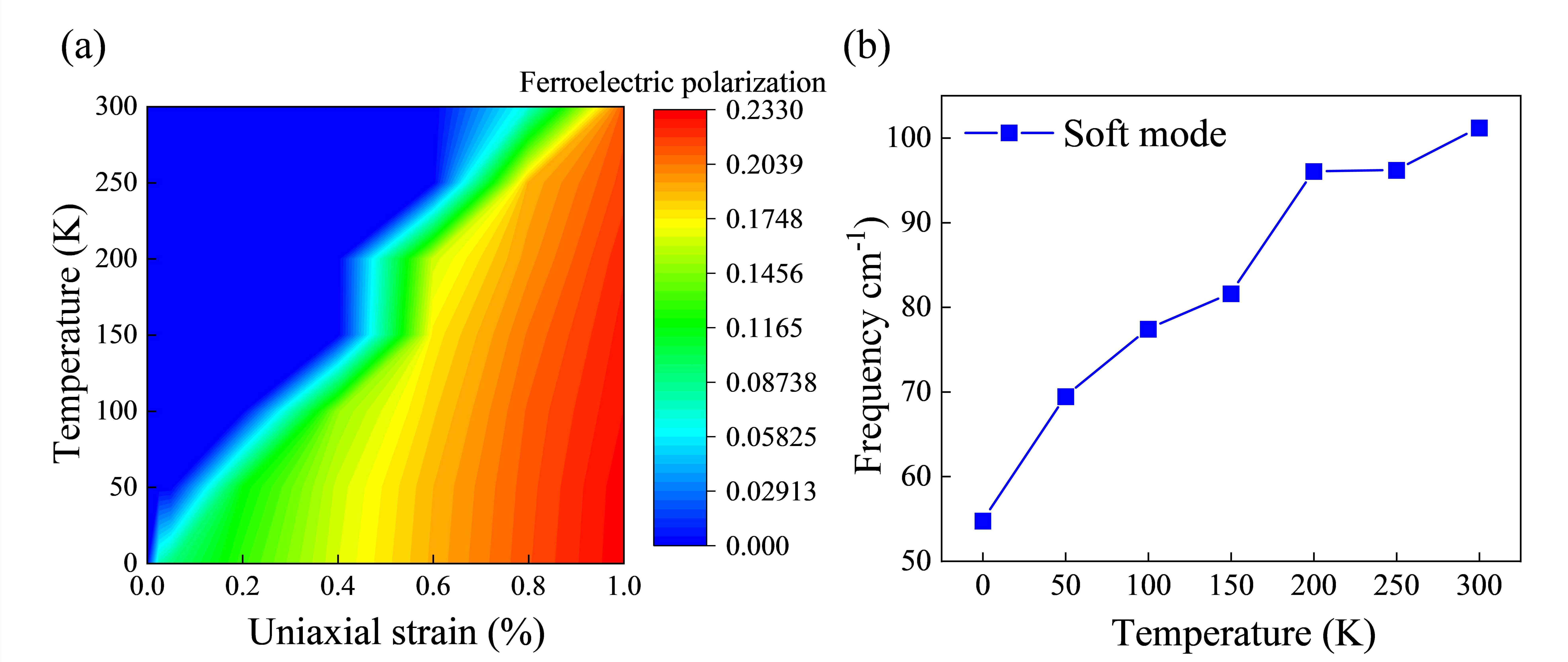}
    \caption{(a) Phase diagram of KTaO\textsubscript{3} under uniaxial strain and at different temperatures and (b) Variation of soft mode in ferroelectric phase 
for 1\% uniaxial strained KTaO\textsubscript{3} across the temperature range of 0-300 K.
}
    \label{fig:hotfigu}
\end{figure}

In the case of strained KTaO\textsubscript{3}, it is crucial to understand the relationship between the ferroelectric energy barrier and quantum fluctuations. We analyze the one dimensional potential energy surface for KTaO\textsubscript{3} under 1\% uniaxial strain at 0 K using the same SCAN functional and computational parameters. The energy barrier in Fig. S2 was found to be $3.1$~meV/atom within DFT calculations, which is comparable to that in BaTiO\textsubscript{3}~\cite{verdi2023quantum}. This result suggests that under strain, quantum fluctuations are no longer sufficient to overcome the energy barrier, leading to the stabilization of the ferroelectric phase.
Additional to the DFT calculation, we also calculated the potential energy surface by MLFF, which is shown by red line in Fig. S2. In comparison, the MLFF method predicts an energy barrier of $3.2$~meV/atom, demonstrating excellent agreement with the DFT result.
\end{spacing}
\vspace{-3mm}
\begin{spacing}{2}
As previously discussed, strained KTaO\textsubscript{3} exhibits significant and systematic structural distortions across various temperatures. To further investigate these distortions, we computed the ferroelectric polarization using the Berry phase method \cite{resta1994macroscopic,king1993theory,solem1993understanding,spaldin2012beginner}. 
The evolution of ferroelectic phase diagrams under different strains and temperatures, presented in   Fig.~\ref{fig:hotfigu} and Fig. S5, is consistent with the bond length trends observed in
Fig.~\ref{fig:bond} and Fig. S1. The exact polarization values for each ferroelectric structure are displayed in Fig. S3 and Fig. S4, with color intensity increasing proportionally to the magnitude. 
In the uniaxial case, the largest ferroelectric polarization occurs at the highest strain, reaching a value of 0.233 $C/m^2$.
Under biaxial strain, the polarizations along both in-plane directions exhibit comparable magnitudes, resulting in a net polarization of 0.293 $C/m^2$ along the diagonal direction. Notably, the maximum ferroelectric polarization in both uniaxially and biaxially strained KTaO\textsubscript{3} is comparable to the polarization observed in BaTiO\textsubscript{3}\cite{king1993theory,rabe1992first,cohen1992origin,merz1954domain}, emphasizing the significant enhancement of ferroelectric properties due to strain. 
\end{spacing}
\vspace{-8pt}
\begin{spacing}{2}
In the ferroelectric phase of strained KTaO\textsubscript{3}, the soft mode frequency remains positive across the entire low-temperature range up to the Curie temperature.
As ferroelectric properties are closely linked to soft phonon modes, and given the approximately linear relationship between ferroelectric polarization and strain, it is reasonable to expect that the soft modes will also exhibit a linear dependence.
To characterize this behavior, we exact the phonon spectrum of KTaO\textsubscript{3} under both 1\% uniaxial and biaxial strain, examining the evolution of soft modes with temperature.
Fig~\ref{fig:hotfigu} (b) and Fig. S6 illustrate the evolution of the soft mode in phonon dispersion for KTaO\textsubscript{3} under 1\% uniaxial and biaxial strain at different temperatures.
From the figures, it can be observed that in both cases, the soft mode frequency shifts to higher values as the temperature increases. This behavior aligns well with the linear trend observed in the ferroelectric polarization.
The results demonstrate that anharmonic effects not only eliminate the imaginary frequencies associated with the ferroelectric soft mode but also exhibit the temperature dependent evolution. 
\end{spacing}

\section{Conclusion}\label{sec4}

\begin{spacing}{2}
In this study, we systematically investigated the effect of in-plane uniaxial and biaxial strain, ranging from 0\% to 1\%, on pristine KTaO\textsubscript{3} to evaluate its potential for ferroelectricity induction via inversion symmetry breaking over the temperature range of 0-300 K. By employing the SSCHA+MLFF workflow and Berry phase method, we obtained accurate structural information and polarization values under different strain conditions while accounting for higher order anharmonic effects.

Our calculations reveal that under uniaxial strain and biaxial strain in the range of 0.2\% to 1\%, the system undergoes a ferroelectric to paraelectric phase transition as temperature increases. Specifically, the space group evolves from P4mm to P4/mmm in the uniaxial strain case and from Amm2 to P4/mmm under biaxial strain. Furthermore, our results demonstrate that, similar to pristine KTaO\textsubscript{3}, the inclusion of anharmonic effects stabilizes the spurious imaginary ferroelectric soft mode predicted by the harmonic approximation.

Additionally, we found that at 0 K, the energy barrier for the system under 1\% uniaxial strain is 3.1 meV/atom, which is comparable to the energy barrier in BaTiO\textsubscript{3} (4 meV/atom), suggesting that under strain, quantum fluctuations are no longer sufficient to overcome the energy barrier, thereby stabilizing the ferroelectric phase in strained KTaO\textsubscript{3}. Using the Berry phase method, we constructed the ferroelectric phase diagram for both uniaxial and biaxial strained cases. Our calculations indicate that the maximum spontaneous polarization occurs at 1\% strain and 0 K, reaching approximately 0.233 $C/m^2$
along a direction for the uniaxial case and 0.293 $C/m^2$ along in-plane diagonal direction for the biaxial case.

By considering both temperature effects and anharmonicity, our study offers theoretical predictions that closely reflect realistic experimental conditions. These findings offer valuable guidance for future experimental and theoretical research on strain-engineered ferroelectricity in KTaO\textsubscript{3}.
\end{spacing}

\section{Methodology}\label{sec3}
\begin{spacing}{2} 

The computational work in this study was conducted by combining Vienna Ab initio Simulation Package (VASP) \cite{kresse1993ab,kresse1996efficiency} with SSCHA method~\cite{hooton1955li,monacelli2021stochastic,bianco2017second,errea2014anharmonic}. DFT calculations were carried out at the meta-GGA level. Specifically, the SCAN functional \cite{sun2015strongly} was employed alongside projector augmented wave potentials \cite{kresse1999ultrasoft}. A plane-wave energy cutoff of 800 eV was used to ensure the accuracy and convergence of the results. The Berry phase theory was conducted to compute the electrical polarization \cite{resta1994macroscopic,king1993theory,solem1993understanding,spaldin2012beginner}.

\begin{figure}[H]
    \centering
    \includegraphics[width=\linewidth]{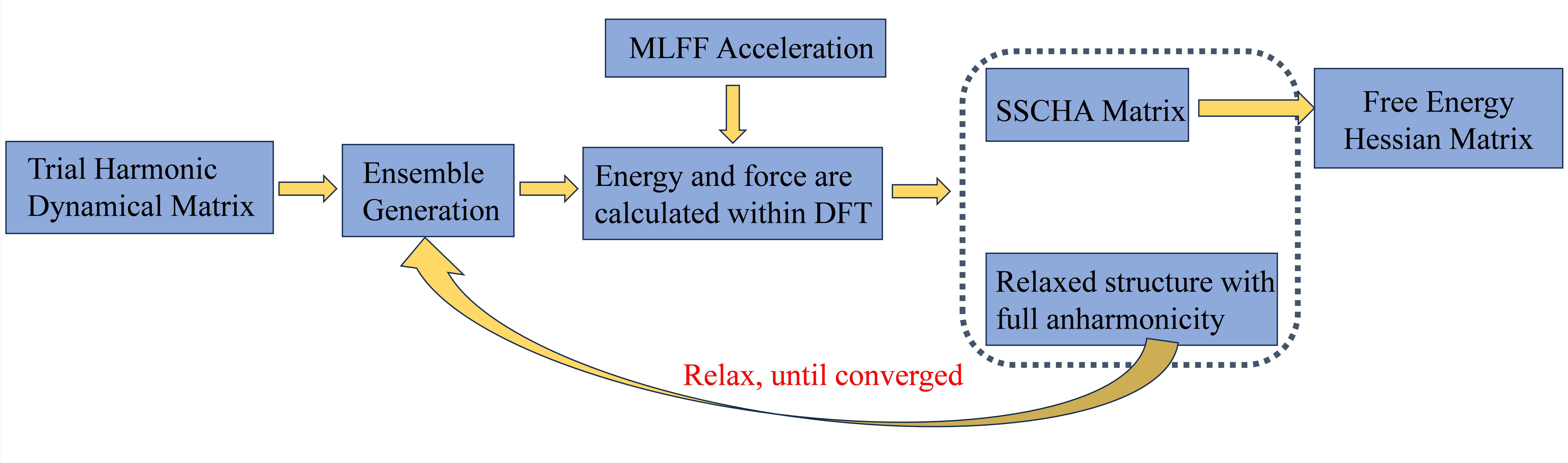}
    \caption{Sketch of the SSCHA+MLFF workflow.}
    \label{fig:workflow}
\end{figure}
\vspace{-15pt}

The prinstine KTaO\textsubscript{3} is modeled using the experimental lattice of 3.9842 Å \cite{samara1973anharmonic,ranalli2023temperature}, and strain is applied on this basis. To prevent the emergence of additional soft ferroelectric modes, only the atomic positions are relaxed, while all lattice constants are kept fixed. To fully consider quantum and anharmonic effects of KTaO\textsubscript{3}, a computational workflow was designed. This workflow integrates a VASP-to-SSCHA interface and incorporates on the fly MLFF \cite{jinnouchi2019fly,jinnouchi2019phase,jinnouchi2020descriptors} to calculate energies and forces for ensembles generated within the SSCHA framework. The structure of this workflow is illustrated schematically in Fig.~\ref{fig:workflow}. The MLFF, trained on the fly from DFT molecular dynamics with Bayesian error estimation, accelerates the stochastic sampling in SSCHA by several orders of magnitude while retaining full fidelity to the ab initio potential energy surface. This integrated framework enables us to compute the temperature dependent ferroelectric soft mode frequency $\mathrm{W}_{\mathrm{FE}}(\mathrm{T})$, anharmonic phonon spectra, and free-energy landscape across a range of strain conditions, thereby mapping the critical strain and temperature thresholds for stabilizing the FE phase. Details of the computational methodology are provided in Supplementary material.
\end{spacing}
\vspace{-1.0em}
\section*{Data availability }

The authors declare that all source data supporting the findings of this study are
available within the article and the Supplementary information file.

\bibliography{sn-bibliography}

\section*{Acknowledgements}
\vspace{-1.0em}
\begin{spacing}{2}
Y.Z gratefully acknowledges the financial support from the Vienna Doctoral School of Physics through fellowship. The work is supported by the Austrian Science Fund (FWF) Grant-DOI: 10.55776/PIN5456724.  The computational results presented
in this article are obtained using the Austrian Scientific
Cluster (ASC).
\end{spacing}

\section*{Author contributions}
\vspace{-1.0em}
\begin{spacing}{2}
Conceptualization:C.F., Y.Z., L.R., and T.C.
Methodology and calculations:
Y.Z., and L.R. Supervision: C.F. Writing: Y.Z., L.R., and C.F.
Discussion and reviewing: All authors.
\end{spacing}
\vspace{-1.2em}
\section*{Competing interests}
The authors declare no competing interests.

\section*{Additional information}
\vspace{-1.0em}
\begin{spacing}{2}
\textbf{Supplementary information}
The online version contains supplementary material available.  

\noindent\textbf{Correspondence}
Correspondence and requests for materials should be addressed to Cesare Franchini and Wei Ren.
\end{spacing}

\end{document}